\documentclass[letterpaper, 10 pt, conference]{ieeeconf}  %
\IEEEoverridecommandlockouts                              

\usepackage{graphics} 
\usepackage{epsfig} 
\usepackage{commath}
\usepackage[notrig]{physics}
\usepackage{stix} 
\usepackage{amsmath,amsfonts}
\usepackage{notoccite}
\usepackage{amsmath} 

\title{\LARGE \bf
Adaptive Drift Compensation for Soft Sensorized Finger Using Continual Learning
}

\author{Nilay Kushawaha$^{1,2}$, Radan Pathan$^{1,2}$, Niccolò Pagliarani $^{1,2}$, Matteo Cianchetti $^{1,2}$ and Egidio Falotico $^{1,2}$
\thanks{*We acknowledge the contribution from the Italian National Recovery and Resilience Plan (NRRP), M4C2, funded by the European Union–NextGenerationEU (Project IR0000011, CUP B51E22000150006, "EBRAINS-Italy")}
\thanks{$^{1}$ The BioRobotics Institute, Scuola Superiore Sant’Anna, Pontedera, Italy}%
\thanks{$^{2}$ Department of Excellence in Robotics and AI, Scuola Superiore Sant’Anna, Pisa, Italy (n.kushawaha, r.pathan, n.pagliarani, m.cianchetti, e.falotico)@santannapisa.it}%
}

\begin{document}

\maketitle
\thispagestyle{empty}
\pagestyle{empty}

\begin{abstract}
Strain sensors are gaining popularity in soft robotics for acquiring tactile data due to their flexibility and ease of integration. Tactile sensing plays a critical role in soft grippers, enabling them to safely interact with unstructured environments and precisely detect object properties. However, a significant challenge with these systems is their high non-linearity, time-varying behavior, and long-term signal drift. In this paper, we introduce a continual learning (CL) approach to model a soft finger equipped with piezoelectric-based strain sensors for proprioception. To tackle the aforementioned challenges, we propose an adaptive CL algorithm that integrates a Long Short-Term Memory (LSTM) network with a memory buffer for rehearsal and includes a regularization term to keep the model’s decision boundary close to the base signal while adapting to time-varying drift. We conduct nine different experiments, resetting the entire setup each time to demonstrate signal drift. We also benchmark our algorithm against two other methods and conduct an ablation study to assess the impact of different components on the overall performance.

\end{abstract}
\section{INTRODUCTION}
Inspired by the morphology of biological organisms, soft robots are designed to be flexible and compliant, enabling them to function effectively in unstructured environments \cite{rus2015design}\cite{laschi2016soft}. In grasping applications \cite{li2024continual}\cite{abondance2020dexterous}, soft grippers with high compliance provide safer interactions with their surroundings and humans, effectively overcoming the limitations of rigid grippers, which struggle to handle delicate objects. Among these, soft grippers equipped with soft pneumatic actuators (SPAs) \cite{pagliarani2023towards} stand out due to their low cost, simple control, and dominance in soft gripper designs. A key perceptive variable for soft grippers is the bending angle, which is crucial in characterizing the bending behavior of soft actuators in feedback control systems. However, the inherent flexibility of soft actuators introduces greater uncertainty in their behaviour, as their configurations can change due to both planned and unplanned deformations \cite{el2020soft}. This unpredictability poses a major challenge for achieving precise sensing and perception in soft actuators.

While incorporating sensing capabilities into soft robots can address some of these challenges, it also introduces new issues such as hysteresis and long-term drift in sensor signals, which can complicate calibration efforts \cite{thuruthel2019soft}. Poor calibration may lead to signal drift, negatively affecting the performance of the control system and limiting its ability to function optimally. These biases can reduce the detection range, and over time result in controller malfunctions. Signal drift in soft sensorized structures arises from the material’s intrinsic properties, such as viscoelasticity, polymer degradation \cite{hutchinson1995physical}, as well as external factors like circuitry noise and the sensitivity of sensors to environmental conditions. This deviation disrupts the mapping relationship, making standard regression methods inadequate and requiring frequent sensor recalibration.

The data-driven approach has gained widespread use in sensor calibration \cite{chin2020machine}\cite{shih2020electronic}. Machine learning techniques, particularly long short-term memory (LSTM) variants, are often employed to perform regression using datasets, without needing to explicitly model the sensor's characteristics \cite{thuruthel2019soft}\cite{kim2018deep}. However, long-term drift can cause significant deviations in the sensor signal from its original state. As drift accumulates over time, conventional data-specific machine learning methods may lose reliability and predictive accuracy. Despite this, recurrent networks are favored because they account for the time-varying dynamics of sensors, which are generally assumed to remain consistent across models. 

Several deep learning techniques have been explored in the literature to address the challenge of sensor signal calibration in the presence of temporal drift. Some researchers have approached the drift issue as a domain adaptation problem, solving it through optimal transportation transfer learning \cite{kim2020adaptive} while others have employed Gaussian mixture domain adaptation \cite{zhang2023gaussian} and deep subdomain learning adaptation networks \cite{zhang2022deep}. Wang et al. \cite{wang2024drift} developed a hybrid architecture, combining an autoencoder (AE) to capture drift-aware temporal relationships from historical data with a long short-term memory (LSTM) network to extract relevant data features and incorporate drift characteristics. Loo et al. \cite{loo2022robust} merged an adaptive unscented Kalman filter with an RNN-based architecture to mitigate sensor noise and drift. Thuruthel et al. \cite{thuruthel2020drift} employed transfer learning to extract latent space representations, allowing knowledge gained from one task to be applied to another for a soft strain sensor embedded in a passive anthropomorphic finger. However, a key limitation of these methods lies in either the complexity of parameter estimation, as in the case of the Kalman filter, or the increased training time and computational cost due to the need to retrain the entire network on newly drifted data. Additionally, all of the above algorithms try to generalize more over the current drift patterns resulting in an increased error when the previous drift patterns repeats itself. Also the use of longer sequences of sensor data, as in \cite{kim2020adaptive}, introduces prediction delays.

In this paper, we investigate the use of CL to address drift in a piezoelectric-based soft sensorized finger for proprioception. CL focuses on progressively training
models on a continuous stream of data to accumulate and retain knowledge over time \cite{kushawaha2024continual}. Inspired by advancements in continual learning, we propose an adaptive training approach that allows a neural network to continuously learn from new drift data while maintaining its performance on the original baseline signal as well as the previous drift data. In addition, our algorithm also demonstrates adequate forward and backward knowledge transfer during training. The effectiveness of the proposed method is validated across nine experimental datasets, with the first serving as the baseline signal and the remaining eight representing drift scenarios. We compare our approach with two benchmark methods and conduct an ablation study to assess the importance of the different parts of the algorithm. To the best of our knowledge, this is the first application of continual learning for drift compensation on a soft sensorized actuator.

\section{EXPERIMENTAL SETUP}
\subsection{Soft Sensorized Finger}
The soft finger utilized in this study is a Pneumatic Network (Pneunet) bending actuator, featuring multiple pneumatic chambers for controlled deformation, based on a design from previous research \cite{pagliarani2023towards}. The finger was fabricated using a standard multi-step molding and casting process. Molds for both the top and bottom layers were 3D-printed with PLA filament on a Raise3D printer. DragonSkin 30A elastomer (Smooth-On) was used as the soft material, mixed in a 1:1 ratio with a small amount of red pigment (Silc Pig, Smooth-On) to enhance visibility.

Traditional commercial sensors are typically rigid, bulky, and stiff, making them unsuitable for soft robots due to their interference with natural motion and flexibility \cite{thuruthel2019soft}. To address these challenges, a flexible piezoelectric sensor, termed Piezola, was incorporated into the bottom layer of the actuator. This thin, wire-like sensor generates a voltage proportional to the actuator's deformation, providing real-time feedback on bending behavior while preserving the flexibility of the actuator. Detailed information on the sensor design and specifications is provided in \cite{yoshida2017flexible}. 
For the fabrication of the sensorized finger, the mold for the bottom layer was modified to include a dedicated channel for securely housing the sensing wire. A 0.3 mm PET layer was selected as a strain-limiting element due to its high tensile stiffness, functioning as a neutral bending plane. The sensor was positioned just above the PET layer, ensuring that when the actuator deforms, it applies controlled tension to the sensor wire. The prepared silicone mixture (DragonSkin 30A) was then poured into the mold, fully embedding both the sensor and PET layer, and cured in an oven at 45°C for 5 hours. This process produced a cohesive bottom layer that integrates both the sensor and strain-limiting layer seamlessly into the actuator structure. Finally, both cured layers (top and bottom) were bonded using uncured silicone material, and an inlet pipe was inserted for pressurization.

The sensor's voltage output was amplified using an operational amplifier (op-amp) circuit in a voltage follower configuration, as described in \cite{kim2022piezoelectric}. To further improve the circuit design, a guard trace was added around the input terminals and other components connected to the op-amp. This guard trace reduces surface charge leakage and minimizes the influence of external noise, ensuring more accurate strain measurement, even for small deformations. The amplified voltage output was then digitized and processed using a data acquisition system (Arduino Due) for further analysis.
\begin{figure}
	\centering
    \includegraphics[]{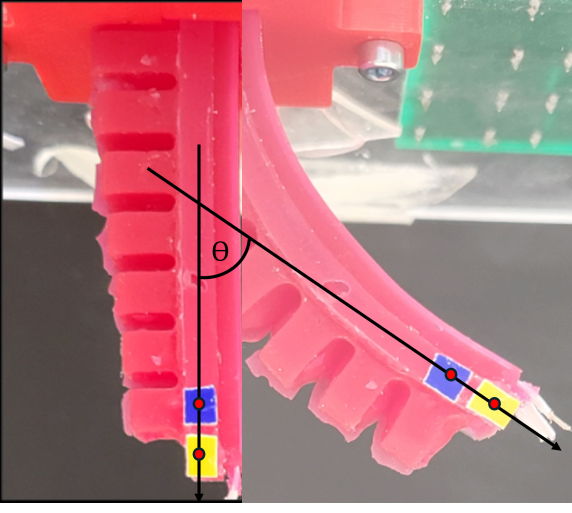}
	\caption{Soft pneumatic finger at rest and actuated state. The tip bending angle is calculated using dot product between the two vectors derived from the vision markers (yelllow, blue).}
	\label{fig:finger_img}
\end{figure}
\begin{figure}
	\centering
    \includegraphics[]{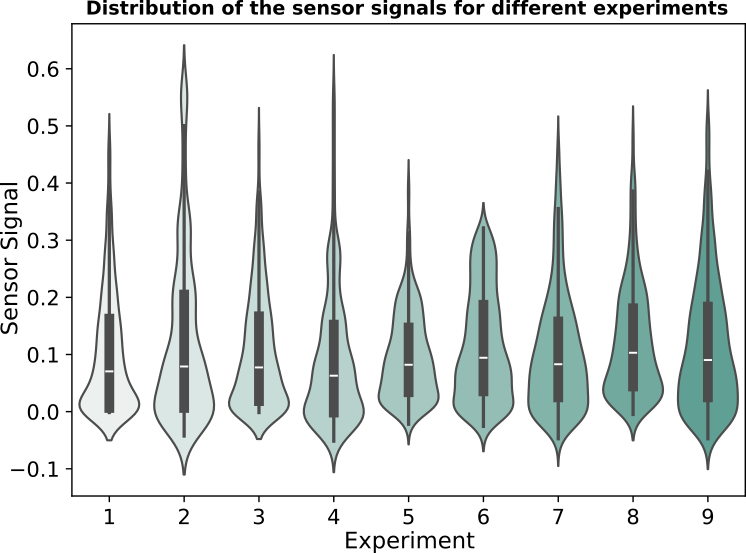}
	\caption{The sensor signals were recorded across nine distinct experiments, with each experiment repeated for five cycles. The complete setup was restarted before each experiment to simulate the drift effects.}
	\label{fig:Sensor signal distribution}
\end{figure}
\subsection{Data Collection}
 To measure the bending angle of the finger, we attach two markers at the finger's tip as illustrated in Figure \ref{fig:finger_img}. We define two vectors: one formed by the two fixed markers and another representing a normal vector, which is computed based on the finger's initial configuration. The angle between these vectors is determined using the dot product formula, as shown in equation \ref{cosine}. 
\begin{equation}
\label{cosine}
    \theta =\cos^{-1}{\left(\frac{\Vec{a}\cdot \Vec{b}}{\norm{\Vec{a}}\norm{\Vec{b}}}\right)}
\end{equation}
Each experiment consists of 5 cycles, reaching a maximum bending angle of approximately 100 degrees, with 1600 data points collected for both the training and testing phases. During each experiment, the setup is restarted, introducing an initial offset in the circuit and creating a virtual drift. Additionally, random drift from the circuit along with environmental factors, as well as the hysteresis of the material contribute to the overall drift in the system. The data distribution of the sensor signal across different experiments is depicted in Figure \ref{fig:Sensor signal distribution}.

\subsection{Task creation}
\begin{figure}
	\centering
    \includegraphics[]{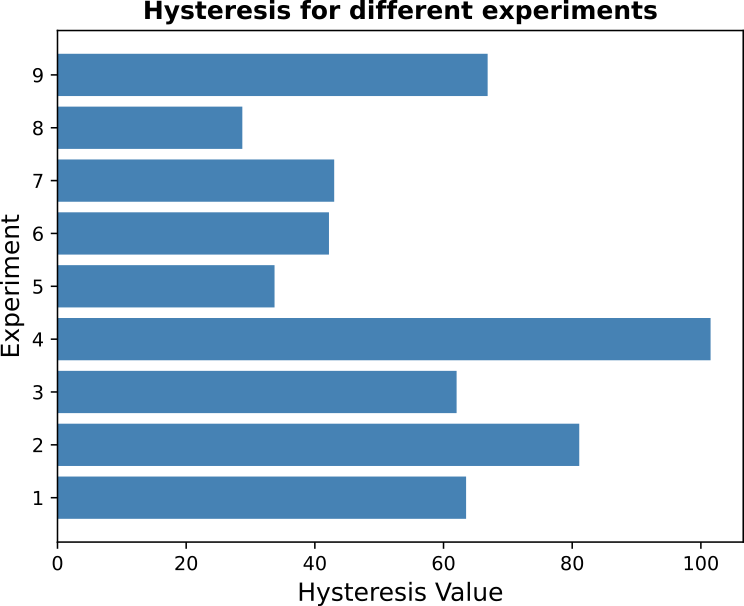}
	\caption{The area under the hysteresis curve (relationship between sensor signal and the bending angle) for different experiments. The experiment with the lowest hysteresis value is selected as the base signal.}
	\label{fig:hysteresis}
\end{figure}

The input features consist of raw sensor signals from the soft pneumatic finger embedded with a piezoelectric sensor. These raw signals, along with the tip bending angle (target variable), are combined to form a structured dataset. The dataset is divided into 9 sub-tasks, where the first task serves as the base signal for all algorithms, while the remaining 8 tasks represent experimental data that exhibit significant drift. The base signal is determined by calculating the area under the hysteresis curve (between sensor signal and tip bending angle) for each experiment and selecting the one with the lowest hysteresis value, as illustrated in Figure \ref{fig:hysteresis}. Experiment 8 was found to have the lowest hysteresis value. The test data is structured in a similar fashion, containing all 9 experiments arranged in the same order as the training dataset.

\section{METHODOLOGY}
\begin{figure}
	\centering
    \includegraphics[]{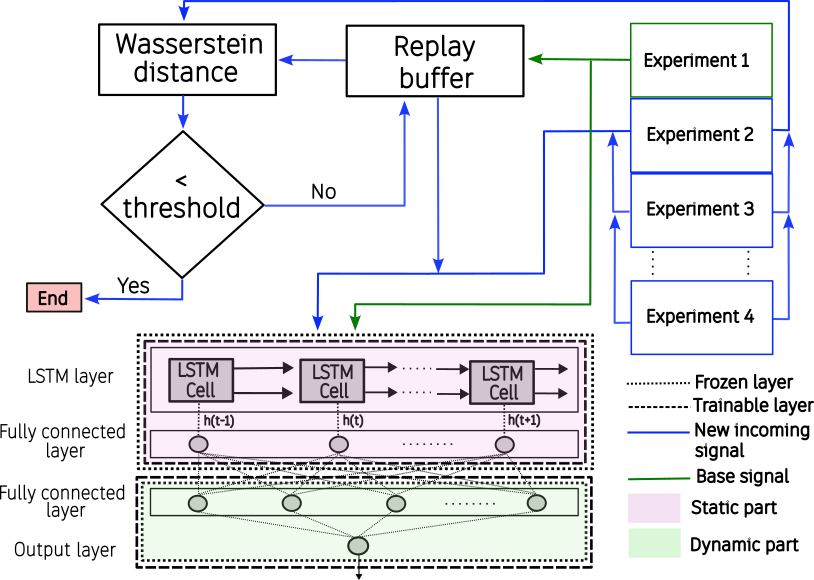}
	\caption{The proposed architecture consists of a static part and a dynamic part with a small replay buffer for rehearsal of the previously learnt knowledge. In addition, a regularization term is added to the loss function to perform knowledge transfer.}
	\label{fig:model_architecture}
\end{figure}
\subsection{Continual Learning Architecture}
The proposed continual learning algorithm employs a hybrid approach \cite{kushawaha2024synapnet}, integrating elements from three traditional continual learning strategies: architectural, regularization, and rehearsal \cite{wang2024comprehensive}. The architectural strategy involves either parameter isolation or a dynamic network architecture. The regularization strategy introduces an additional regularization term to the loss function, which constrains the updates of the neural weights. The rehearsal strategy stores a subset of previously learned tasks in a memory buffer and replays them during training on new tasks.

The network is structured into two distinct parts. The first part, referred to as the static part, consists of a single LSTM layer with 16 nodes, followed by a flattening layer and a fully connected layer with 64 output neurons. The second part, called the dynamic part, comprises two fully connected layers with 64 and 256 neurons, respectively, and a single output neuron. ReLU activation is applied after each linear layer, except for the final output layer. Both sections of the network are trained using mean squared error (MSE) as the loss criterion. Each part employs its own optimizer, the static part uses the Adam optimizer with default settings, while the dynamic part also incorporates a weight decay term $(1e^{-4})$ to regularize the training process and prevent overfitting during adaptive learning.

\subsection{Model Training}
The input to the LSTM network is a raw, unbatched 1D sensor signal, and the final output consists of the corresponding bending angles in degrees. The static part of the network is initially trained on the base signal with the lowest hysteresis value, while the weights of dynamic part remains frozen with the weights initialized to one in the beginning. Once the initial training is completed, the weights of the static part are fixed. For subsequent training, incoming data from new experiments are first compared to previously encountered signal distributions by calculating the Wasserstein distance between the incoming signal and those stored in the replay buffer. If the Wasserstein distance exceeds a predefined threshold, the algorithm enters an adaptive training phase. Otherwise, the network proceeds to the test phase, where the input is processed by the entire network, and the output layer predicts the bending angle.

During the adaptive training phase, the static part of the network is unfrozen and briefly retrained on the new signal, while the dynamic part remains frozen. After this short training session, the static part is frozen again, and the dynamic part becomes trainable. The dynamic part is incrementally trained on the new signal, utilizing a combination of rehearsal and regularization strategies to prevent catastrophic forgetting of prior knowledge. A rehearsal buffer stores one cycle information for each incoming signal and replays it to the dynamic model during training. Additionally, a regression version of Learning Without Forgetting (LWF) \cite{li2017learning} regularization term is added to the loss function, encouraging the network to transfer knowledge from previous tasks to new ones by minimizing the L1 loss between them. This combination of techniques helps the model to maintain the decision boundary close to the previously learned knowledge (base signal) while also allowing it to acquire new information (from drift data).
Adaptive training typically takes around 75 seconds on an Intel Core i7 7th gen CPU $@$ 2.80 GHz, equipped with an Nvidia GTX 1080 GPU, running Ubuntu 20.04.

\section{EVALUATION CRITERIA}
\subsection{Metrics}
To evaluate the performance of the different methods we use the root mean squared error (RMSE) as the primary metric. We calculate the RMSE of the model after it has been trained incrementally on all the experiments data, unless otherwise stated. This allows us to understand how much the model retains the previously learned knowledge after training on new tasks. We also calculate the $R^2$ score of the different models which highlights the capability of the model to make predictions based on the raw sensor data.  In addition to this we also calculate some more CL metrics like the backward knowledge transfer, forward knowledge transfer and forgetting \cite{jeeveswaran2023birt} given by equation \ref{bwt}, \ref{fwt}, and \ref{forget}. Since most of these metrics are defined for classification problems in terms of accuracy so we redefine them by leveraging RMSE as the base metric:
backward knowledge transfer (BWT) refers to the influence of the learning new task $t$ on previously seen tasks $k < t$. Lower BWT implies that the learning task $t$ decreased the error on previous tasks, while a higher value indicates that the learning task $t$ negatively affected the performance of the model on previous tasks. Similarly, forward knowledge transfer (FWT) denotes the effect of learning task $t$ on the future tasks $k > t$. A smaller FWT value denotes the positive impact of learning current tasks on the future tasks. Forgetting refers to the difference between the maximum accuracy of the model in previously learned tasks throughout the learning process and the accuracy after learning the current task. It quantifies the decrease in the performance of previous tasks on learning new tasks.
\begin{equation}
    \label{bwt}
    BWT = \frac{1}{T-1}\sum_{j=1}^{T-1}a_{T,j} - a_{j,j}
\end{equation}
\begin{equation}
    \label{fwt}
    FWT = \frac{1}{T-1}\sum_{j=2}^{T}a_{j-1,j} - a^{'}_{j}
\end{equation}
\begin{equation}
\label{forget}
    Forgetting = \max_{\l\in {1,2...,k-1}} a_{k,l} - a_{k,T} 
\end{equation}
where $a^{'}_{j}$ is the RMSE of the task $j$ at some random initialization, $a_{k,\_}$ refers to the accuracy of the model for task k. All the metrics are calculated after the end of training on the respective tasks. A point to note is that it is ambiguous to calculate the BWT for the first task and the FWT for the last task.
\begin{table*}[]
\caption{RMSE score of the proposed algorithm and the two benchmark algorithms for nine different experiments}
\label{pred_table}
\begin{center}
\begin{tabular}{|c||c||c||c||c||c||c||c||c||c|}
\hline
Model type & Exp0 & Exp1 & Exp2 & Exp3 & Exp4 & Exp5 & Exp6 & Exp7 & Exp8\\[1mm] 
\hline
Baseline model & \textbf{1.67}$_{\pm 0.45}$& 9.85$_{\pm 0.49}$ & 17.79$_{\pm 2.19}$ & 8.53$_{\pm 0.53}$ & 11.07$_{\pm 0.82}$ & 5.49$_{\pm 0.34}$ & 6.75$_{\pm 0.30}$ & 10.46$_{\pm 0.97}$ & 8.43$_{\pm 0.86}$\\[1mm] 
\hline
TL model & 5.35$_{\pm 1.74}$ & 8.19$_{\pm 1.67}$ & 12.30$_{\pm 2.15}$ & 7.19$_{\pm 1.62}$ & 9.72$_{\pm 1.98}$ & 6.38$_{\pm 1.44}$ & 5.62$_{\pm 1.53}$ & 8.79$_{\pm 1.70}$ & 7.23$_{\pm 1.93}$ \\[1mm]
\hline
CL model & 6.43$_{\pm 2.02}$ & \textbf{3.48}$_{\pm 1.85}$ & \textbf{3.045}$_{\pm 4.31}$ & \textbf{2.48}$_{\pm 1.90}$ & \textbf{2.86}$_{\pm 1.83}$ & \textbf{3.59}$_{\pm 1.52}$ & \textbf{2.96}$_{\pm 2.28}$ & \textbf{2.62}$_{\pm 1.66}$ & \textbf{2.07}$_{\pm 2.03}$\\[1mm]
\hline
\end{tabular}
\end{center}
\end{table*}
\subsection{Benchmark Models}
We evaluate the performance of our adaptive algorithm by comparing it with a traditional baseline model and a transfer learning approach proposed by Thuruthel et al. \cite{thuruthel2020drift}. All the benchmark algorithms use the same architecture and training parameters as our proposed method. The baseline model merges both the static and dynamic components into a unified single network. In contrast, the transfer learning algorithm initially trains the static component on the first task (base signal), then freezes it, and trains only the adaptive component on the new experiment signals. The key distinction between their approach and ours lies in the inclusion of an adaptive training rule and the integration of CL strategies to enhance the training process, allowing for better generalization across both old and new tasks.
\section{RESULTS}
\begin{figure}
	\centering
    \includegraphics[]{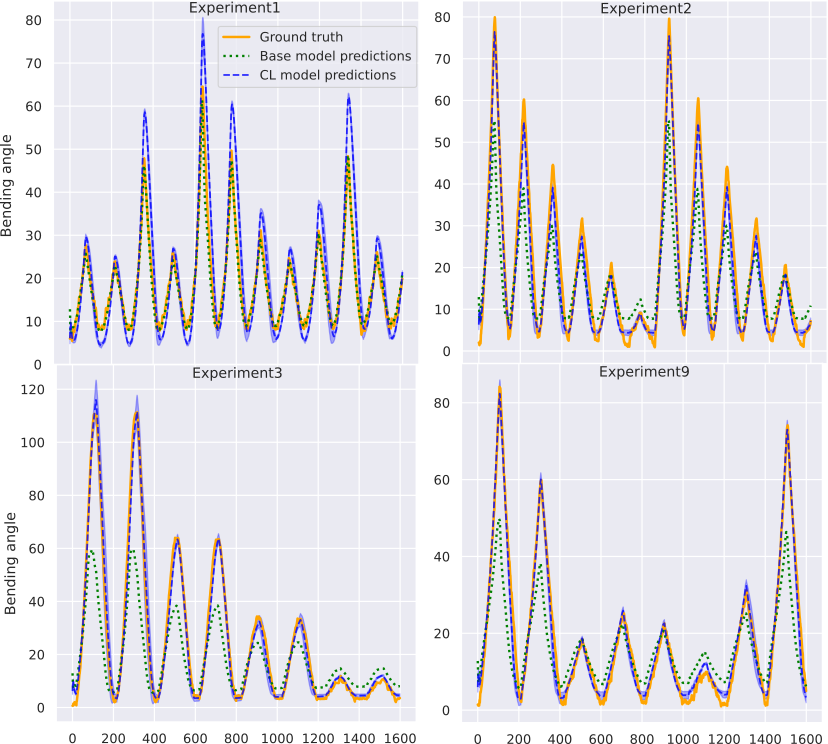}
	\caption{The figure shows the predicted bending angles for different experiments. On top-left, the first experiment which we consider as the base signal/experiment is plotted. In this particular case the performance of the baseline model is better than the proposed CL model.}
	\label{fig:model_pred}
\end{figure}
The performance of the proposed algorithm is assessed through various tests, focusing primarily on evaluating the RMSE error in its predictions across both previous as well as the current tasks. 
Additionally, the study calculates forward knowledge transfer, backward knowledge transfer, and forgetting of the proposed method. Furthermore, we also report the R$^2$ score for all the 3 algorithms and perform an ablation study. All experiments are repeated five times, and we plot the mean and standard deviation of the respective metrics. The terms "tasks" and "experiments" are used interchangeably throughout this analysis.
\begin{figure}
	\centering
    \includegraphics[]{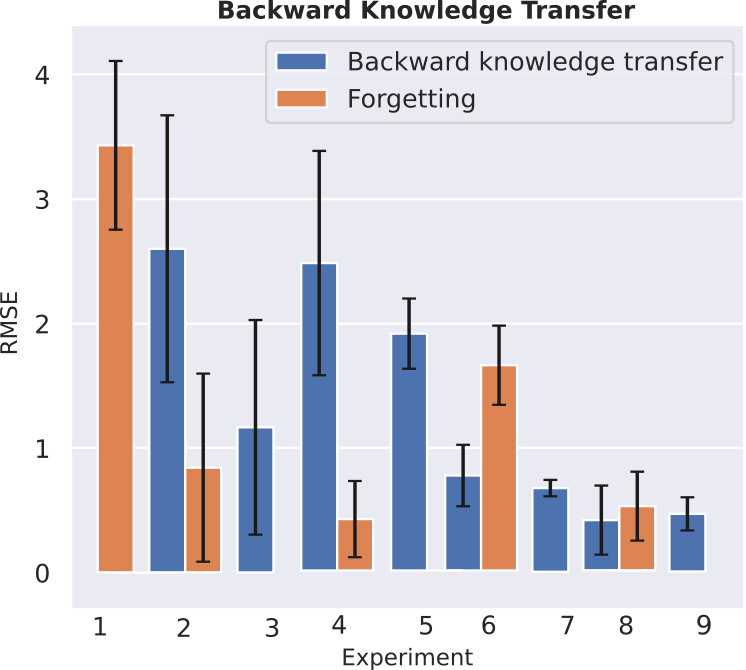}
	\caption{BWT and forgetting of the proposed CL algorithm, forgetting is calculated after the model has been trained on all the tasks, where as the BWT is calculated at the end of each task.}
	\label{fig:bwt}
\end{figure}
Figure \ref{fig:model_pred} shows the predictions of the CL model (in blue solid line) as well as the baseline model (in green dashed line) on the test signal. An important point to note is that the CL model predictions are calculated once the model has been trained on all the experiments data. It is evident that the CL model is able to consistently maintain a lower error on the previous as well as the new experiments unlike the baseline model that performs optimally on the initial task, but experiences a decline in performance on subsequent tasks. Table \ref{pred_table} shows the performance of the different algorithms across all the nine experimental datasets. The CL algorithm surpasses the performance of both the benchmark algorithms in every task, except for the first, where the baseline model achieves the highest performance.

\begin{figure}
	\centering
    \includegraphics[]{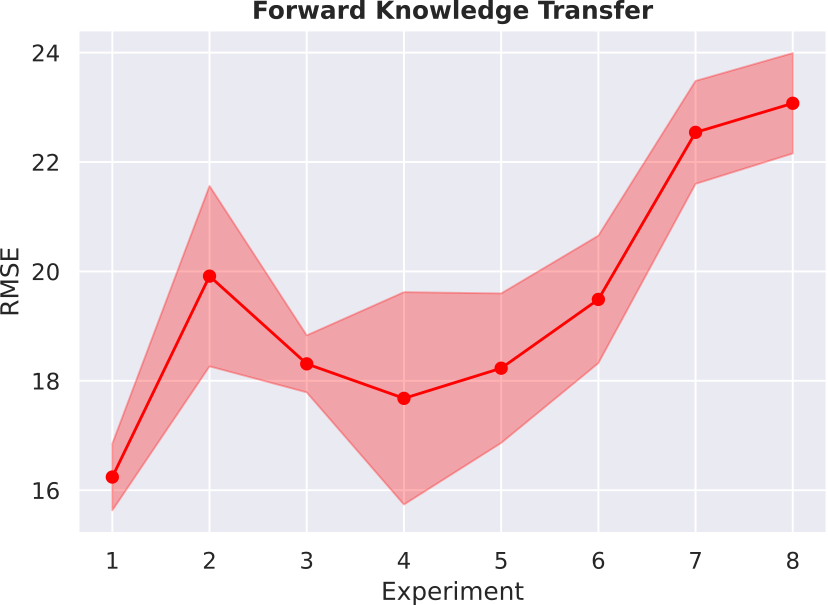}
	\caption{FWT for the different experiments, similar to BWT it is also calculated at the end of each task.}
	\label{fig:fwt}
\end{figure}
The forgetting and backward knowledge transfer observed in the various experiments are illustrated in Figure \ref{fig:bwt}. Notably, forgetting remains nearly negligible for some tasks, which can be attributed to the similarity between those tasks and the final task. In contrast, backward knowledge transfer becomes more pronounced after training on more recent tasks, as the model retains a greater amount of information for generalization. As previously noted, backward knowledge transfer is zero for the first task, while forgetting is zero for the last task. Conversely, forward knowledge transfer is most significant for the first task, reflected by a lower RMSE value, which tends to increase for subsequent tasks. This trend may be explained by the need for highly specific knowledge in later tasks, whereas for the first task, even a general understanding can enhance the RMSE of downstream tasks. Overall, these experiments confirm that the base signal remains consistent across all experiments, with only drift being added to the data.
\begin{figure}
	\centering
    \includegraphics[]{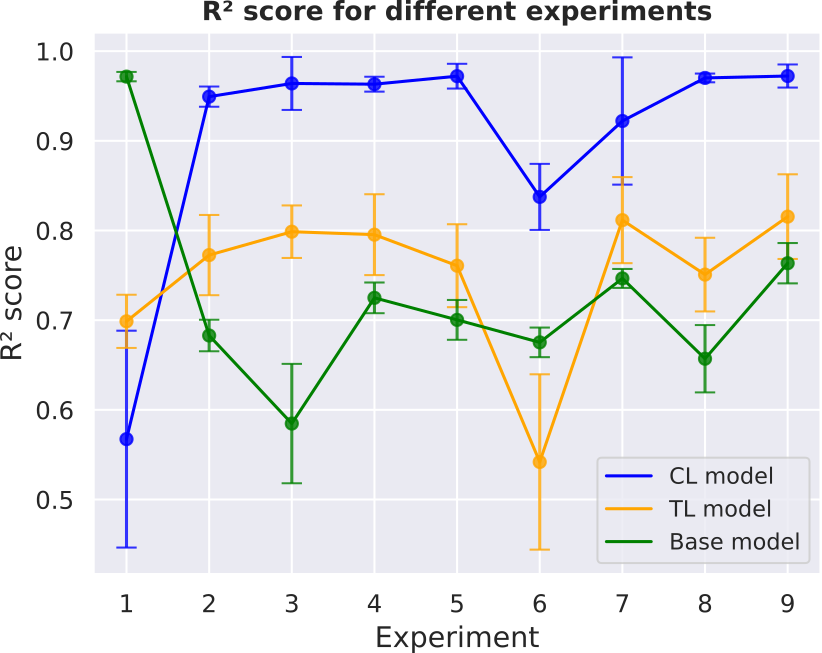}
	\caption{The CL model maintains a high R$^2$ value across all experiments except the first one, as it tries to generalize across most tasks rather than focusing solely on the initial task like the base model.}
	\label{fig:R_sq score}
\end{figure}
The R$^2$ score measures the goodness of fit of a regression model by indicating how well the model's predictions match the actual data (the higher the value, the better the fit). As shown in Figure \ref{fig:R_sq score}, the CL model achieves the best performance for all the experiments except for the first task where the base model and the TL model outperforms it. The TL model performs better than the base model except for the first and sixth task.
\begin{figure}
	\centering
    \includegraphics[]{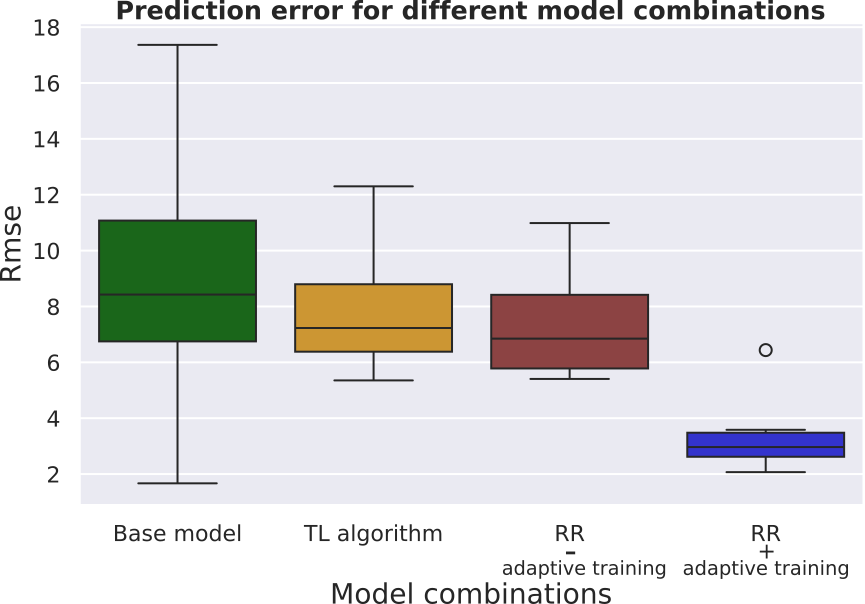}
	\caption{The overall RMSE score for the base model, TL model and the CL model without and without the adaptive training phase.}
	\label{fig:ablation}
\end{figure}
Figure \ref{fig:ablation} depicts the average overall RMSE of the different algorithms for all the tasks. The base model has the highest error followed by the TL algorithm, "Regularization \& Rehearsal (RR) - adaptive training" refers to the case where the complete model (combining the static part and the dynamic part) is trained every time on the new experiment data incrementally, with both replay and regularization tricks to avoid catastrophic forgetting but without the adaptive training phase. It performs a little better than the TL algorithm, however the maximum performance boost is attained after incorporating the adaptive training phase in addition to the CL tricks represented by "RR + adaptive training".

\section{CONCLUSIONS}
This work presents an adaptive continual learning algorithm for modelling the behaviour of a soft sensorized finger to compensate for the overall system drift. The algorithm is initially trained on the base signal to capture its temporal patterns. When drift occurs in subsequent experiments, the algorithm enters an adaptive training phase, incrementally learning from the drift data over a short period. To prevent catastrophic forgetting of previously acquired knowledge, we utilize a combination of architectural modifications, regularization techniques, and rehearsal strategies. The algorithm's performance is evaluated using data collected from a soft finger equipped with an embedded piezoelectric strain sensor. In addition to learning new information from the drift data, the algorithm also demonstrates effective forward and backward knowledge transfer. We compare our method to a transfer learning algorithm from \cite{thuruthel2020drift} and a naive model trained only on the initial base signal. 
Although training on the entire dataset at once would mitigate the drift, this is impractical due to the unavailability of the complete data at once as well as extended training time and computational resources.

However, one key limitation of our approach is the need to store a subset of prior experimental data in a replay memory for rehearsal. In future work, we aim to eliminate the replay buffer by using an autoencoder-based generative model, which would serve dual purposes: functioning as a replay buffer for rehearsal and providing a mechanism to detect when drift occurs.

\addtolength{\textheight}{-12cm}  
\bibliographystyle{ieeetr}
\bibliography{references_file.bib}

\end{document}